\documentclass{aa}
\usepackage{graphicx}
\usepackage{txfonts}
\usepackage{float,psfig}
\usepackage{aalongtable}
\begin{document}

\title{Suzaku observations of X-ray excess emission in the cluster of galaxies A\,3112}

\author{T. Lehto\inst{1}
          \and
          J. Nevalainen \inst{1,2}
          \and
          M. Bonamente\inst{3}
          \and
          N. Ota\inst{4}
          \and
          J. Kaastra \inst{5,6}
          }

   \institute{Division of Geophysics and Astronomy, Department of Physics, University of Helsinki, Finland\\
              \email{tuomas.kz.lehto@helsinki.fi}
         \and
             Finnish Centre for  Astronomy with ESO, University of Turku, V\"ais\"al\"antie 20,
             FI-21500 Piikki\"o, Finland
         \and
             University of Alabama in Huntsville, USA\\
         \and
             Department of Physics, Tokyo University of Science, 1-3 Kagurazaka, Shinjuku, Tokyo 162-8601, Japan
         \and
             SRON Netherlands Institute for Space Research, Sorbonnelaan 2, 3584 CA Utrecht, the Netherlands \\
         \and       
             Sterrenkundig Instituut, Universiteit Utrecht, P.O. Box 80000, 3508 TA Utrecht, The Netherlands\\
              }

   \date{Received; accepted}

\abstract
{}
{We analysed the Suzaku XIS1 data of the A\,3112 cluster of galaxies in order to examine the X-ray excess emission in 
this cluster reported earlier with the XMM-Newton and Chandra satellites.}   
{We performed X-ray spectroscopy on the data of a single large region. We carried out simulations to estimate the
systematic uncertainties affecting the X-ray excess signal.} 
{The best-fit temperature of the intracluster gas depends strongly on the choice of the energy band used for the spectral
analysis. This proves the existence of excess emission component in addition to the single-temperature MEKAL in A\,3112. 
We showed that this effect is not an 
artifact due to uncertainties of the background modeling, instrument calibration or the amount of Galactic absorption. 
Neither does the PSF scatter of the emission from the cool core nor the projection of the cool gas in the cluster 
outskirts produce the effect. Finally we modeled the excess emission either by using an additional MEKAL or powerlaw 
component. Due to the small differencies between thermal and non-thermal model we can not rule out the non-thermal origin of the 
excess emission based on the goodness of the fit. Assuming that it has a thermal origin, we
further examined the Differential Emission Measure (DEM) models. We utilised two different DEM models,
a Gaussian differential emission measure distribution (GDEM) and WDEM model, where the emission measure of a number of
thermal components is distributed as a truncated power law. The best-fit XIS1 MEKAL temperature for the 0.4--7.0 keV band is
4.7$\pm0.1$ keV, consistent with that obtained using GDEM and WDEM models.}   
{}


\authorrunning{T. Lehto et al.}
\titlerunning{X-ray excess in A\,3112 cluster with Suzaku}

\maketitle

\section{Introduction}
Several studies have found that while clusters of galaxies emit X-rays via bremsstrahlung and line emission, some of them 
also possess an additional X-ray emission component (see e.g. Durret et al. 2008). This emission has been usually dubbed 
as soft excess since if assuming that the cluster emission in the highest energies (E $\ge$ 2 keV) is correctly modeled 
with a thermal model, extrapolation of this model yields excess emission at the soft X-rays. However, the waveband where 
the excess emission dominates depends on the adopted energy range for the fitting of the thermal 
model. In this Paper, we call this effect X-ray excess.  

Since the typical X-ray excess signal level is only  $\sim$10\% of that of the main cluster emission, and at the level of 
calibration uncertainties, it is difficult to characterise in detail. Usually thermal and non-thermal models yield equally 
good fits to the data. However, the existence of such an additional component may still bias the best-fit temperatures if 
not accounted for. Furthermore, due to its different spectral shape, the additional emission component will contaminate 
the main thermal emission component by a varying amount in different bands. This will result to different temperature 
values when fitting the contaminated spectrum in different wavelength bands with a thermal model. We will use this effect 
to examine the existence of the X-ray excess emission in the A\,3112 cluster. The X-ray excess emission is previously observed in A\,3112 
with XMM-Newton (Nevalainen et al., 2004) and Chandra (Bonamente et al., 2007). A\,3112 is the BM type I cluster and its richness class is 2.

A possible wide-band systematic error on the calibration of a given instrument may in principle yield different temperatures when fitting 
different bands, even if the real emission was thermal emission with a single temperature. 
Indeed, Nevalainen et al. (2007) showed that the improved calibration of the XMM-Newton EPIC instruments between 
2002 and 2005 led to significant changes in the amount of X-ray excess emission in several clusters.
However, if several independently calibrated instruments show consistently the same effect, it is unlikely to be due to the 
calibration problems. In this paper, we will address this issue by examining the data of the cluster A\,3112 obtained with 
several independent instruments (Suzaku XIS unit 1; XMM-Newton PN, MOS1 and MOS2).

Another crucial factor in the X-ray excess signal analysis is the amount of Galactic absorption. The long standing 
standard for the hydrogen column density NH has been the 21 cm measurements by Dickey \& Lockman (1990). Recently, this 
study has been superseded by the Leiden-Argentine-Bonn study (Kalberla et al., 2005).
In the case of A\,3112, the two works yield significantly different results for the column density. 
The new value, 1.3 $\times 10^{20}$ cm$^{-2}$ is only $\sim$50\% of the previous value.
Other things being the same, the new, lower value of NH will reduce the amount of soft X-ray excess from that derived 
using the old NH  (as in Nevalainen et al. 2004; Bonamente et al. 2007). We will examine in this paper the effect of  
NH on the X-ray excess emission in A\,3112.

Throughout the paper we use the following cosmological model, $H_0$=70 km/s/Mpc, $\Omega_m$=0.3 and $\Omega_{\Lambda}$=0.7. 
The redshift of the cluster is 0.0753. Physical size of the studied 3--6' region corresponds to 0.26--0.51 Mpc at cluster redshift.

\section{Suzaku data processing}
\label{processing}
We describe here the basic steps in the Suzaku data processing (for the XMM-Newton data processing, see Appendix 
\ref{xmm_app}).
We observed A\,3112 with Suzaku in May 2008. The ID of the observation is 803054010.
We used only the data from the X-ray Imaging Spectrometer (XIS) unit 1, since it has better response at lower energies than
units 0 and 3 (Koyama et al. , 2007).
The XIS was operated in the normal clocking mode. The 3x3 and 5x5 editing modes were combined totalling 57 ks of exposure time. The calibration files were those
available on 2010 January 22. The event screening was started from the unfiltered event files and we reprocessed the data using the latest
calibration files along with HEASOFT version 6.8. 
The standard selection criteria were applied to the data concerning passage through the South Atlantic Anomaly, 
geomagnetic cut-off-rigidity $>$~6~GV, and elevation angles less than 5$^{\circ}$ and 20$^{\circ}$ from the 
nighttime and daytime Earth. 
In addition, the data at the detector regions containing the calibration sources were 
rejected. The full FOV XIS1 light curves in soft (E$<$2 keV) and hard band (E$>$2 keV) showed that the variation 
of the count rate is within $\pm$10\% and rather consistent within the statistical uncertainties. Thus, there was no need
to filter the data further for rejecting particle flare periods (see also Section \ref{trans_app}). 

The energy Redistribution Matrix Files (RMF) and the Ancillary Response Files (ARF) were created using the 
{\tt xisrmfgen} and {\tt xissimarfgen} tools, respectively. ARFs were created separately for the cluster emission and for 
the sky background. As the cluster brightness distribution model, we used a simulated image based on the best-fit surface 
brightness profile model obtained from XMM-Newton PN data (see Section \ref{psf_app}). 
For the sky background we used a uniform brightness distribution.
 
We created the particle-induced instrumental background (NXB) spectra using the {\tt xisnxbgen} tool version from 2008 March 8, 
which uses collected night Earth data. The tool uses the revised geomagnetic cut-off-rigidity (COR2) as the NXB indicator
(see Tawa et al., 2008). The generated NXB spectra were subsequently subtracted from the cluster data.

\section{Spectral analysis issues}
\label{bkg}
\subsection{General}
Unless stated otherwise, we modeled the cluster emission spectra using a single temperature MEKAL model,
absorbed by the WABS model with NH = 1.3 $ \times 10^{20}$ cm$^{-2}$ (Kalberla et al., 2005). 
We used the abundance table of Anders \& Grevesse (1989) and the XSPEC version 12.5.1n. 
We excluded photons with energy above 7 keV and below 0.4 keV due to background (see below) and to minimise the XIS 
contamination problems. We binned the spectra to contain a minimum of 100 counts per bin. 
The emission from the central region is complicated due to cooling and the presence of a bright
AGN (Takizawa et al., 2003), while the outer region is problematic due to relatively stronger background.
Thus, in this work we analysed only the annular region at a distance of 3--6 arcmin from the cluster center. 

\subsection{Non-X-ray background}
\begin{figure}
\resizebox{\hsize}{!}{\includegraphics{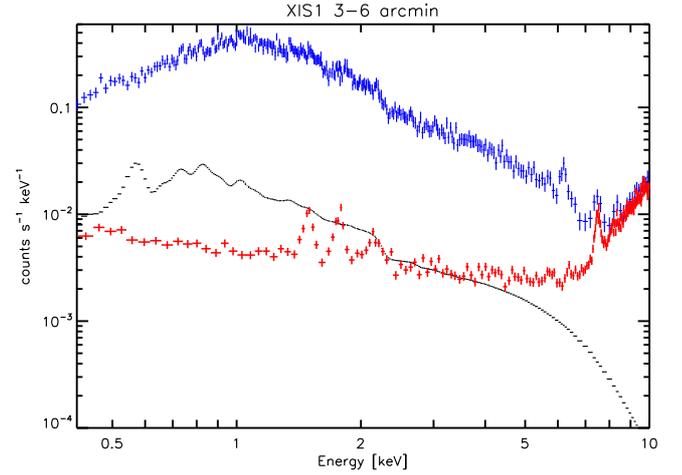}}
\caption{The total emission (blue crosses), non-X-ray background (red crosses) and the sky background (black curve) in the 3--6 arcmin region for XIS1.
\label{bkg.fig}}
\end{figure}

The Non-X-ray-background emission, NXB, is below 10\% of the cluster signal at the low end of our energy range
(0.4 keV). Towards the highest energies, the cluster signal decreases following the decreasing effective area while the NXB 
emission remains relatively constant, until it rapidly increases above photon energies of 7 keV (see Fig. \ref{bkg.fig}). 
At the highest photon energies (7.0 keV) in our adopted range, 
the NXB reaches a level of 30\%  of the cluster signal. Above 7 keV the instrumental Ni line emission exceeds the cluster signal.

Tawa et al. (2008) examined the reproducibility of the non-X-ray background modeling using several different methods.
The one relevant to the current public version of the NXB modeling (the usage of the anti-correlation of NXB with the 
revised cut-off-rigidity COR2) reproduces the NXB data in the 1--7 keV band for a 50 ks exposure 
within 2--3 \% accuracy. Thus, the maximal relative effect in the background-subtracted signal in any of our channels 
due to NXB modeling uncertainty is less than 1\%, i.e. negligible.

\subsection{Sky background}
\label{sky}
A single on-axis pointing does not allow an estimate of the local background.
We used the Rosat All Sky Survey based tool of the HEASARC web page to extract the sky background from the distance of 0.2-1.0$^{\circ}$ 
of the A\,3112 center.
We modeled the spectrum with a MEKAL + WABS(MEKAL + POW) model where the first MEKAL component is due to the Local Hot Bubble (T=0.12 keV),
and the second one (T=0.42 keV) models the Galactic Halo. The abundances were kept in the solar values. POW is the power-law 
component modeling the Cosmic X-ray Background. We kept 
the photon index fixed to 1.41 as found in the measurements by De Luca et al. (2004).

In the spectral analysis, we multiplied the background model by the auxiliary response file produced assuming an uniform 
spatial distribution of the emission, relevant to the sky background. We scaled this model according to the size of the 
regions used to extract the A\,3112 spectra and also considered the source\_ratio\_reg parameter which has been factored into
the xissimarfgen-generated auxiliary response file (Ishisaki et al. 2008). The thus produced background emission prediction was
added to the cluster emission prediction (which was produced using a separate auxiliary response file, see above) and the 
total emission was compared to the data in the fitting process.

In the 0.4--7.0 keV band in the 3--6 arcmin region of A\,3112 the sky background remains below 10\% of the cluster signal (see Fig. \ref{bkg.fig}).
Allowing a conservative 10\% variability for the background due to statistical uncertainties of the RASS measurements yields a negligible 1\% effect on the 
background-subtracted signal.

\subsection{Point sources}
The relatively large PSF of Suzaku causes the random point sources in the line of sight and the AGN embedded in the 
cluster to be confused with the cluster emission. Using the better spatially resolved data from the XMM-Newton satellite (see below) we estimated 
that the point source flux in the 3--6 arcmin annulus is below 0.1\% of the cluster flux, i.e.
negligible.

\section{Results of the spectral analysis}

\subsection{Lower energy band cut-off effect}
\label{cutoff}
In order to examine the possible existence of an additional emission component, we first fitted the 3--6 arcmin spectrum 
with upper energy band cut-off at 7 keV and varying the lower energy cut-off from 2.5 keV to 0.4 keV. 
We fitted the data from the XIS1 separately and the data from the PN, MOS1 and MOS2 simultaneously.

\begin{figure}
\resizebox{\hsize}{!}{\includegraphics{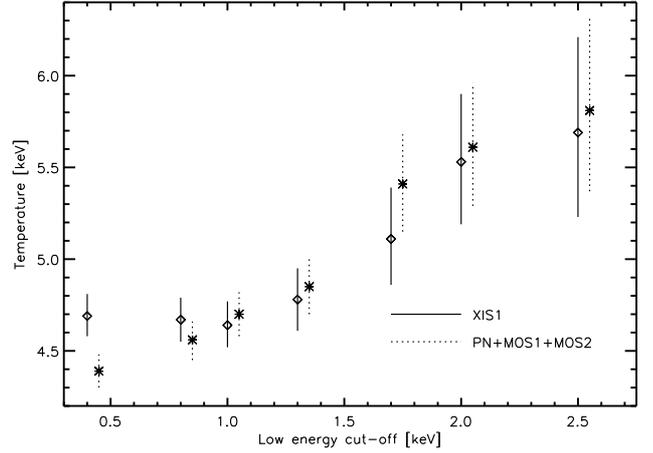}}
\caption{The best-fit temperature and statistical uncertainties at the 1 $\sigma$ level as a function of the lower energy 
cutoff to the XIS data (diamonds and solid line) and XMM-Newton data (asterisks and dotted line) for the 3--6 arcmin region. 
The XMM-Newton values are slightly off-set in the x-direction for clarity. 
\label{cutofftest.fig}}
\end{figure}

The analysis revealed that the best-fit temperature decreases systematically with the lower cut-off value (see Fig. \ref{cutofftest.fig}). 
The best-fit temperatures obtained with a cut-off energy below 1.7 keV are significantly
lower than those obtained with a higher energy cut-off. Using a lower energy cut-off at 2.5 keV or 0.4 keV
yields a best-fit Suzaku temperature of 5.7 keV and 4.7 keV, respectively. 
Since both the Suzaku and XMM-Newton data exhibit a qualitatively similar effect, it cannot be due to calibration problems, 
unless one would assume that all the instruments would have a similar bias in the calibration.
Rather, this effect proves evidence for excess emission in A\,3112 in addition to that of a single temperature MEKAL model.
We also examined the effect of independent abundances for different elements using VMEKAL
model but that did not affect the results.

\begin{figure}
\resizebox{\hsize}{!}{\includegraphics{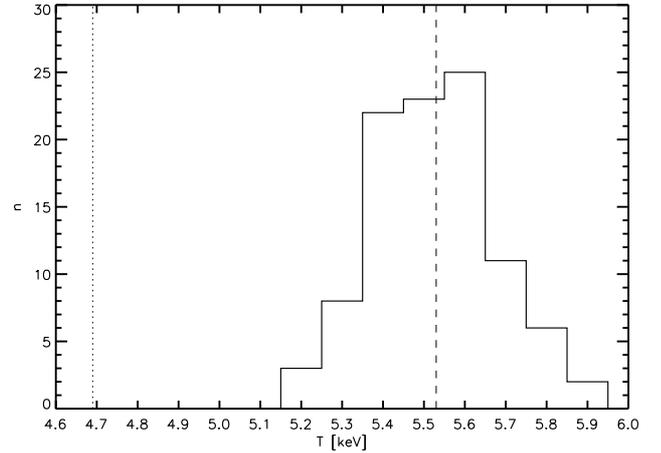}}
\caption{The distribution of the temperature obtained by spectral fits to the simulated data using a cut-off energy of 0.4 
         keV. The simulations were carried out with T=5.5 keV (dashed line). The dotted line shows the best-fit temperature 
         of the observational data using a cut-off energy of 0.4 keV.
\label{T_0.4_histo}}
\end{figure}

If the number of source photons in a given observation is too low, the statistical uncertainties may produce a decreasing 
trend of best-fit temperatures with low energy cut-off due to random effects. As can be seen in Fig. \ref{cutofftest.fig}, 
the effect is not a random fluctuation at the 68\% confidence level. To examine this more, we
simulated 100 XIS1 spectra with the actual exposure time and responses used in our analysis. 
The input model was a single-temperature MEKAL model as obtained from the $E>$2 keV band fit.
We then fitted the simulated spectra similarly as the observational data above, i.e. with a varying low energy cut-off. 
We found that the distribution of the temperature values obtained
by the spectral fits to the simulated data using a cut-off energy of 0.4 keV indicates no
bias towards lower temperatures (see Fig. \ref{T_0.4_histo}).
Only 5 of the simulated sets yielded systematically decreasing trend of temperatures,
when going towards lower cut-off energy, which increases the confidence level of the
measured trend to 95\%.

\begin{figure*}
\centering
\includegraphics[width=9cm, angle=0]{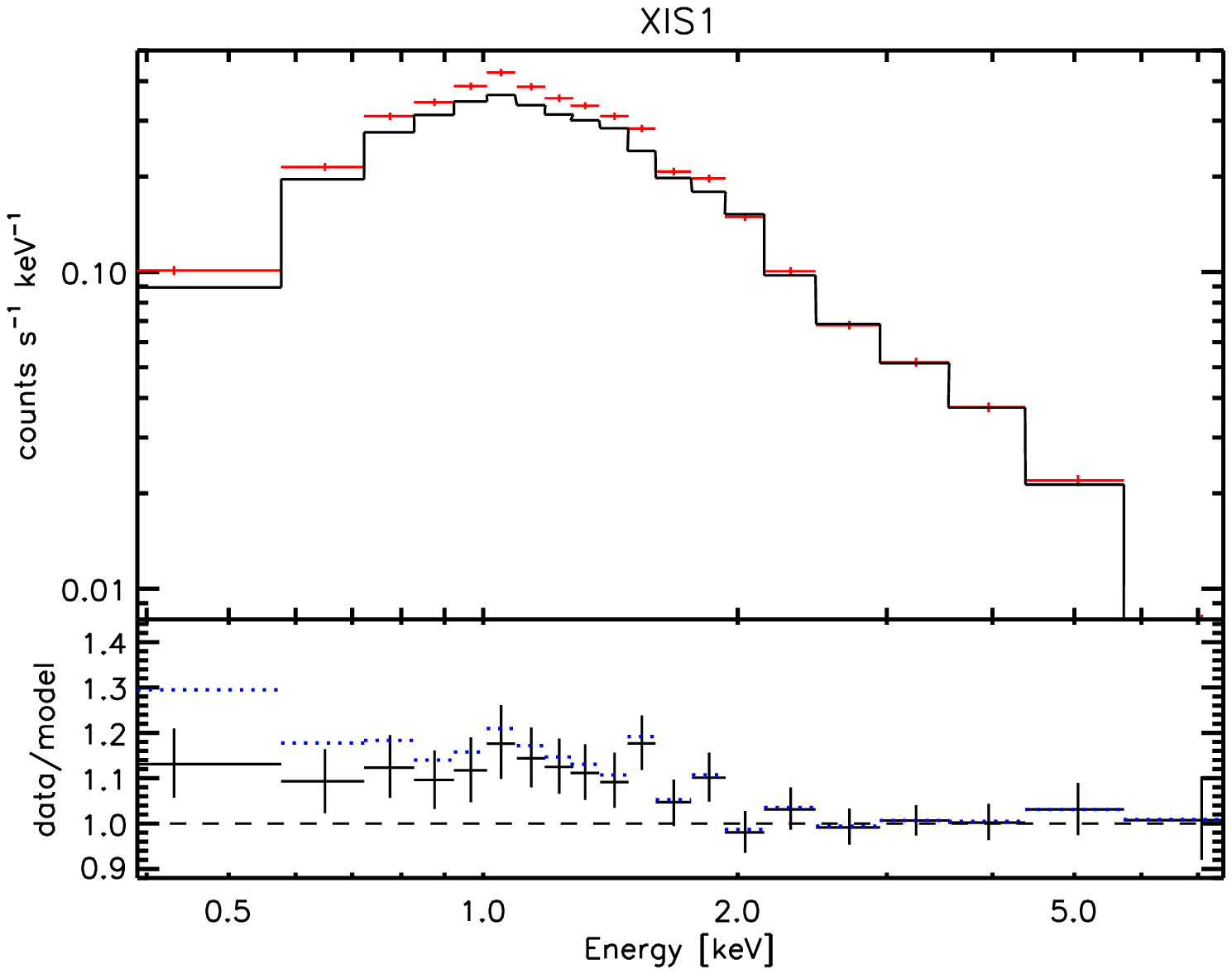}\includegraphics[width=9cm, angle=0]{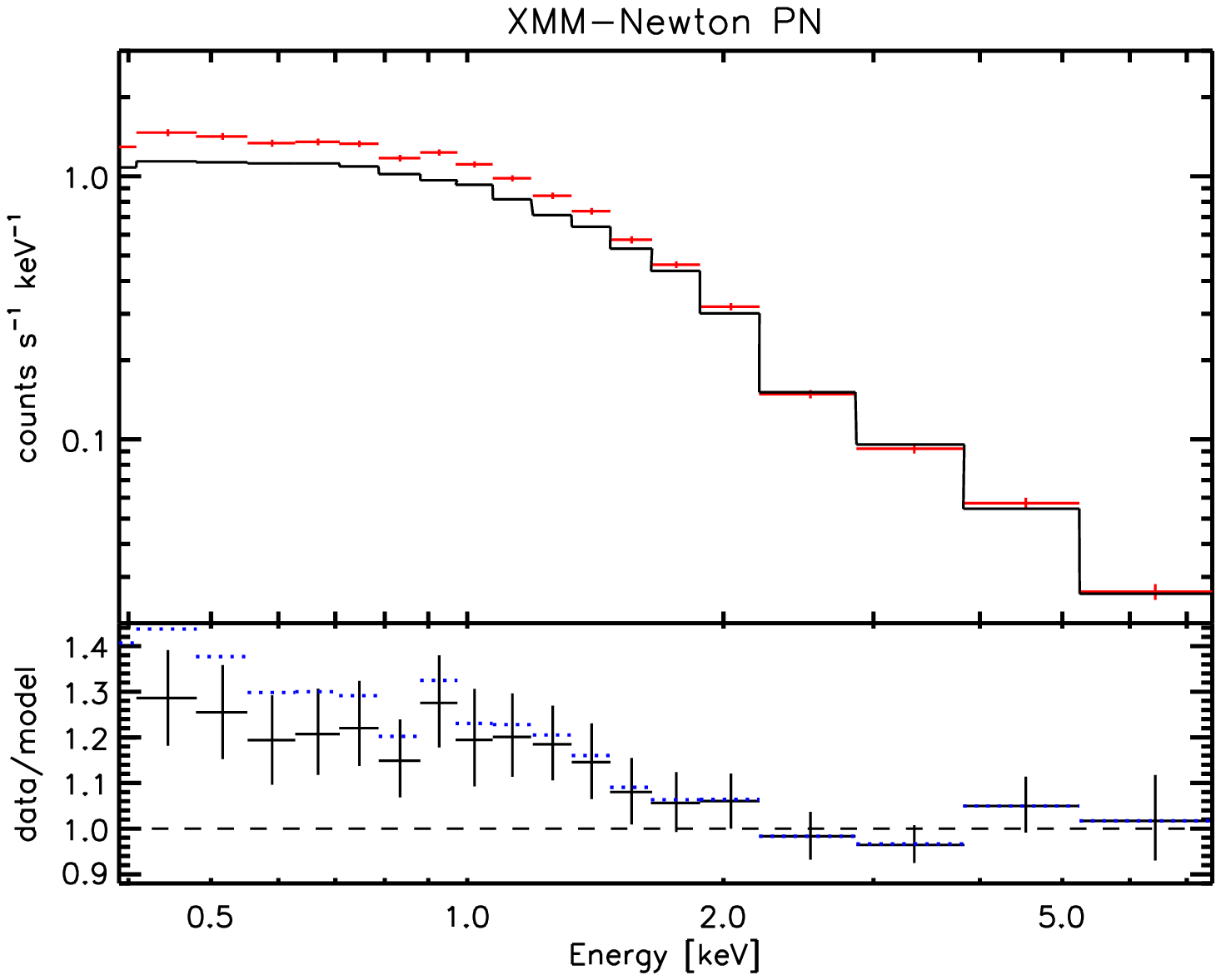}
\caption{The background-subtracted data (red crosses) of the 3--6 arcmin annulus are shown with the thermal model 
prediction extrapolated from the 2--7 keV band (solid black line) in the upper panels. The data are binned using a 
minimum of 2000 counts per bin for clarity. The ratio of the data and the 
extrapolated model is shown with black crosses 
in the lower panels. The error bars include the effect of the statistical uncertainties on the hard band model.
The dotted blue line shows the results of using NH from Dickey \& Lockman (1990). The results are shown for  
XIS1 (left panel) and XMM-Newton PN (right panel). 
\label{extrap.fig}}
\end{figure*}

Assuming that the hard band ($E>$ 2 keV) best-fit single-temperature model is a correct description of the cluster 
emission in that band, we extrapolated it to lower energies. This reveals the excess emission in the XIS1 at
a $\sim$10\% level of the cluster hot gas emission in the 0.4--1.7 keV band (see Fig. \ref{extrap.fig}). 
For comparison, we performed a similar analysis to the XMM -Newton data (see Fig. \ref{extrap.fig}). PN indicates a $\sim$20\% level
for the excess emission. However, the excess emissions in XIS and PN are consistent 
within the statistical uncertainties. 
Using the old Galactic NH value (Dickey \& Lockman, 1990) the excess emission level is twice as high as in the Kalberla et al. (2005)
at 0.6 keV and almost thrice as high at 0.4 keV,
while at higher energies the effect is not significant compared to the statistical uncertainties.

\subsection{Wide band}
\begin{figure}
\resizebox{\hsize}{!}{\includegraphics{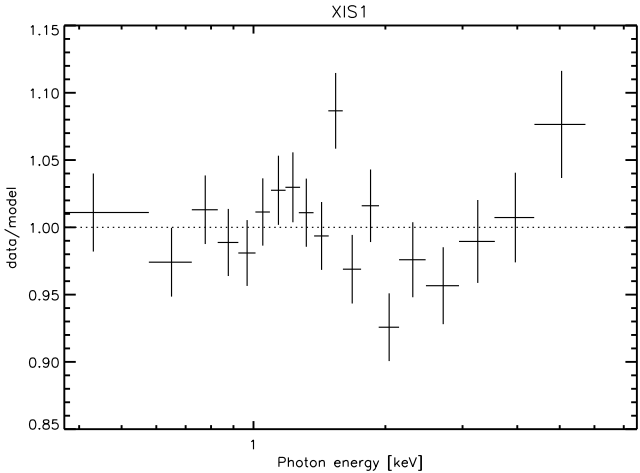}}
\caption{The ratio of the background-subtracted XIS1 data in the 3--6 arcmin annulus to the best-fit single temperature MEKAL model prediction, 
         as fitted to the 0.4-7.0 keV band. The data is binned by a minimum of 2000 counts per bin for clarity.
\label{singleT.fig}}
\end{figure}

The X-ray excess emission should also manifest itself as residuals when the wide band emission is fitted with a 
single-temperature thermal model with the best-fit temperature of 4.7$\pm0.1$ keV. The residuals of the fit are 
at 10\% level (see Fig. \ref{singleT.fig}).
However, the residuals exhibit a systematic trend of increasing values between energies 2 and 7 keV,
which is not consistent with the current knowledge of the calibration accuracy.
The residual peak around 1 keV is consistent with PSF scatter from the high metal abundance core (see Section 
\ref{psf_app}).

\section{Systematic uncertainties}
We examined here whether the above dependence of the best-fit temperature on the choice of the energy band is due to any
of the known sources for systematic uncertainties in the analysis.

\subsection{Contamination of XIS instruments}
Repeated observations of E0102-72.3 by the Suzaku team have yielded information on the accuracy of the time-dependent 
modeling of the contamination in the XIS detectors.
The June 2008 measurements\footnote{http://heasarc.gsfc.nasa.gov/docs/heasarc/caldb/suzaku/docs/ xis/xiscontami\_20090212\_memo.pdf},
which were done less than two weeks after our observation,
indicate that the contaminate model fits the data well within the statistical uncertainties. 

However, we further investigated the effect of the contamination on XIS1 by using the statistical uncertainties of N$_{C} 
\sim 1\times 10^{17}$~cm$^{-2}$ from the memo$^1$. We then examined the effect of additional absorption by multiplying our basic spectral model 
by the VPHABS model, 
where we set the amount of other elements to zero but used the number densities given by the assumed composition of the contaminate 
(C$_{24}$H$_{38}$O$_{4}$) for N$_{\rm C}$, N$_{\rm H}$ and N$_{\rm O}$. 
Due to the restrictions of modeling we assume that the contaminate has a symmetric effect, i.e. a magnitude of the effect when the basic spectral model
is adjusted by reducing or adding the absorption is the same.

\begin{figure}
\resizebox{\hsize}{!}{\includegraphics{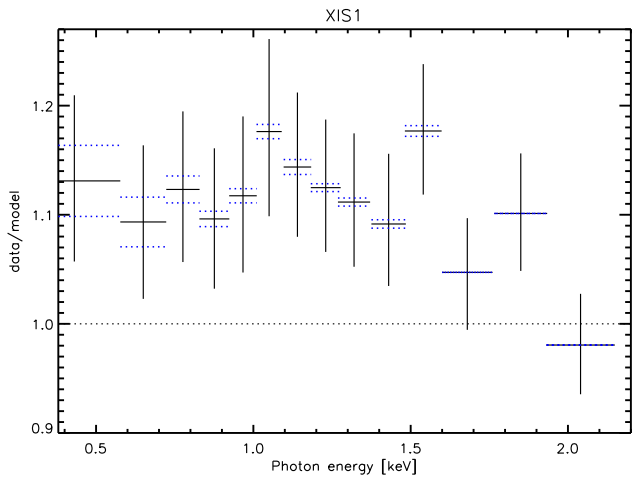}}
\caption{The black crosses are repeated from Fig. \ref{extrap.fig} (the ratio of the data and the best-fit model for XIS1). 
         The blue dotted line shows the ratio when the model is adjusted by reducing or adding the absorption due to uncertainties 
         in the modeling of the contamination (see text). The black dotted line shows the data/model 1:1 ratio.  
\label{xis1_extrap2.fig}}
\end{figure}

The analysis showed that changing the absorption by the above model, the best-fit temperatures with different cut-off 
energies do not change significantly. Extrapolating the $E>$ 2 keV band best-fit model to lower energies with the adjusted 
absorption shows that the contamination effect is negligible compared to statistical uncertainties (see Fig. \ref{xis1_extrap2.fig}).

\subsection{Charge transfer contamination}
\label{trans_app}
Geocoronal and heliospheric Solar wind charge exchange (Wargelin et al., 2004; Snowden et al., 2004) is a possible source 
for contamination of the cluster emission. It produces emission lines in the 0.5--0.9 keV band whose total flux varies by an 
order of magnitude at a timescale of a few months, reaching a level comparable to that of the 
sky background (Galactic emission + CRB) at maximum. Since the sky background 
flux in the 0.4--0.9 keV band is $\sim$10\% of the cluster signal level in the A\,3112 observation, the charge transfer can 
produce an effect that is at the 10\% level of the cluster emission, i.e. comparable to the soft excess signal, provided that
the charge transfer flux is at its maximum during our observation.

\begin{figure}
\resizebox{\hsize}{!}{\includegraphics{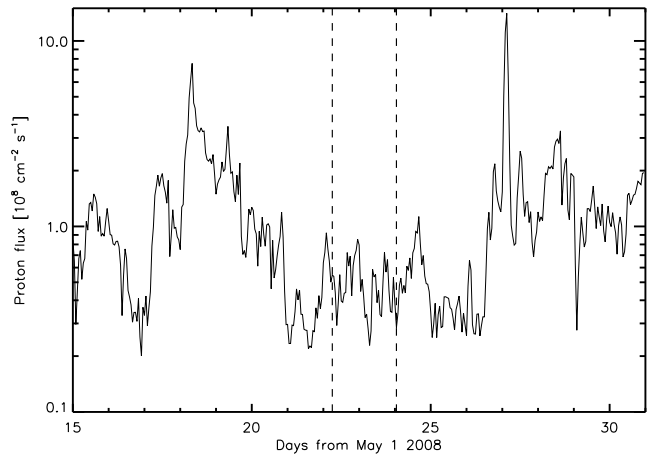}}
\caption{The proton density light curve as measured by the SWEPAM instrument onboard the ACE satellite (curve) and the 
start and stop times of the A\,3112 Suzaku observation (dashed vertical lines).
\label{protons.fig}}
\end{figure}

We examined the ACE SWEPAM solar particle data\footnote{http://www.swpc.noaa.gov/ftpmenu/lists/ace2.html}
of May 2008 in order to find out the Solar activity level during the Suzaku A\,3112 observation. 
The light curve shows that while the proton flux varies by a factor of $\sim$100 during May 15-31, 2008, 
its flux is close to the minimum of the period during the Suzaku observation (see Fig.  \ref{protons.fig}).
Thus, the charge transfer - induced flux remains at a 1\% level of the cluster signal, i.e. negligible.

\subsection{Point spread function}
\label{psf_app}
The relatively large Suzaku point spread function (half power diameter of $\sim$110 arcsec) 
complicates the analysis since the true cluster flux originating from a given region may be contaminated by the 
PSF-scattered flux from neighbouring regions (e.g. Sato et al., 2007; Reiprich et al., 2008).
The effect of the PSF-scattered flux to the analysis of a given cluster region depends on
e.g. the surface brightness and temperature distribution of the cluster, the studied region's size and distance from 
the cluster center, and the cluster redshift. Thus the PSF scatter effect needs to be estimated case-by-case.

To minimise the PSF scatter to and from our studied region of A\,3112, we used a 3 arcmin width for the annular region 
that we studied in this work, i.e. 1.6 times the half power diameter of the PSF.
To minimise the scatter from the cool center we excluded the central r=3 arcmin from our analysis. 

For the PSF scatter estimation, we first fitted the XMM-Newton PN surface brightness data of A\,3112 with a double-beta 
model (Cavaliere \& Fusco-Femiano, 1976), obtaining best-fit values of r$_{core,1}$ = 1.10$\pm$0.03 arcmin and r$_{core,2}$ = 0.22$\pm$0.01 arcmin for the
core radii and  I$_{0,2}$/I$_{0,1}$ = 6.5$\pm$0.4 for the central brightness ratio of the two $\beta$-model components,  
and the common slope parameter $\beta$ = 0.64$\pm$0.01.
Chandra analysis of A\,3112 revealed that the cooling is confined within the central 70 arcsec (Takizawa et al., 2003),
which is accompanied by the central increase of the metal abundance. 
The temperature in the central 0.5 arcmin region is $\sim$ 35\% lower than that at the distance of 3 arcmin from the 
center. Adopting this radial behaviour, and our $E > 2$ keV band fit value of kT $\approx$ 5.5 keV at the 3--6 arcmin 
annulus, we adopt a central temperature of 3.5 keV.

Next we performed ray-tracing simulations with {\tt xissim} (version 2009-01-08) to
estimate the fraction of photons originating from the central cool region into
the 3--6 arcmin annulus. The effect of the attitude drift was taken
into account by incorporating the attitude file in the
simulation. Since the size of the cool emission region is small compared to
the typical size of the Suzaku PSF, we modeled the surface brightness
distribution of the cool component as being constant within 0.5 arcmin
from the cluster center. We assumed the input spectrum to be
monochromatic and with E = 1~keV because the energy dependence of the PSF is
known to be small for the present purpose (see also the Appendix of Sato
et al. 2007).

\begin{figure*}[ht]
\centering
\includegraphics[width=9cm]{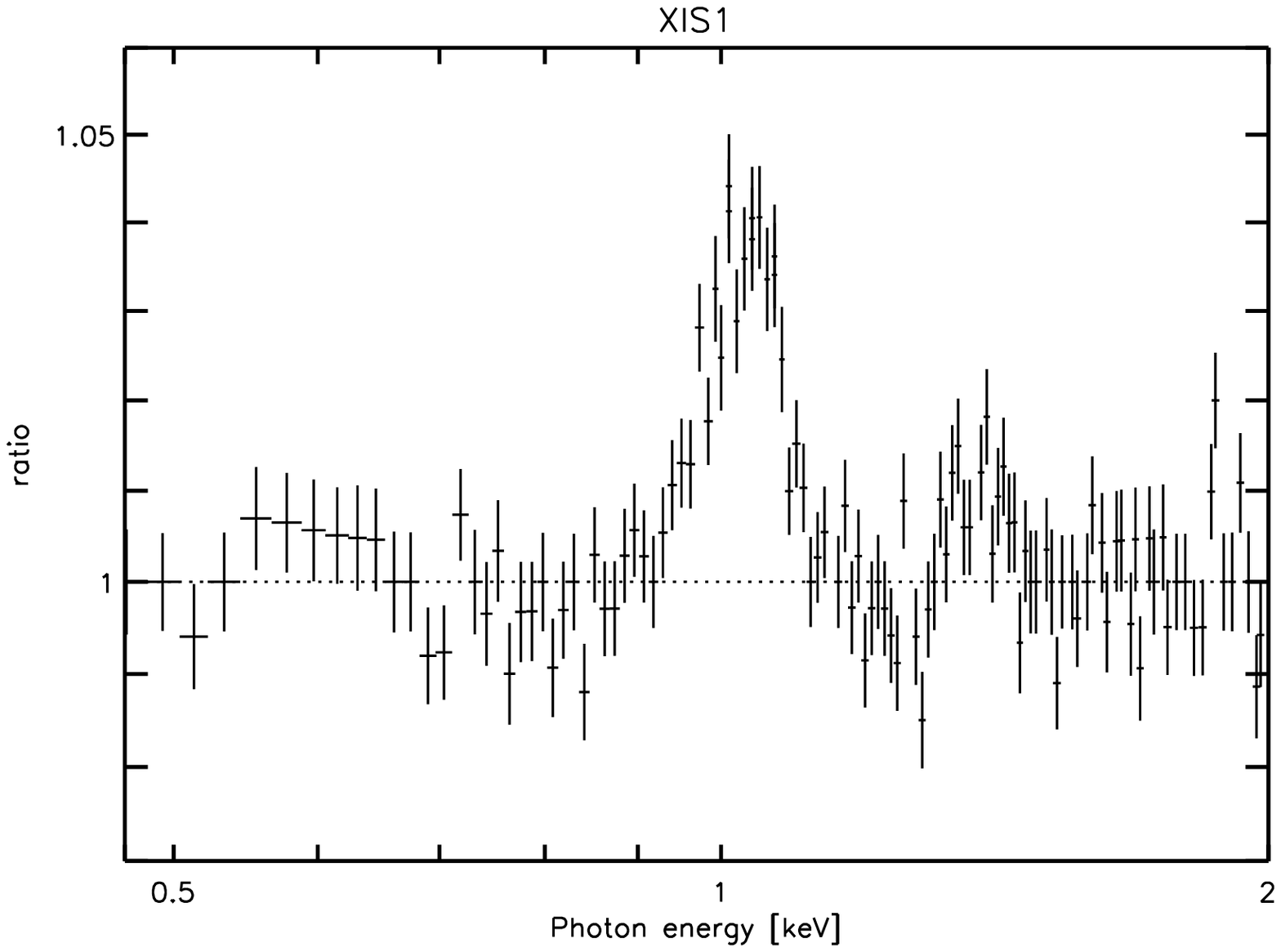}\includegraphics[width=9cm]{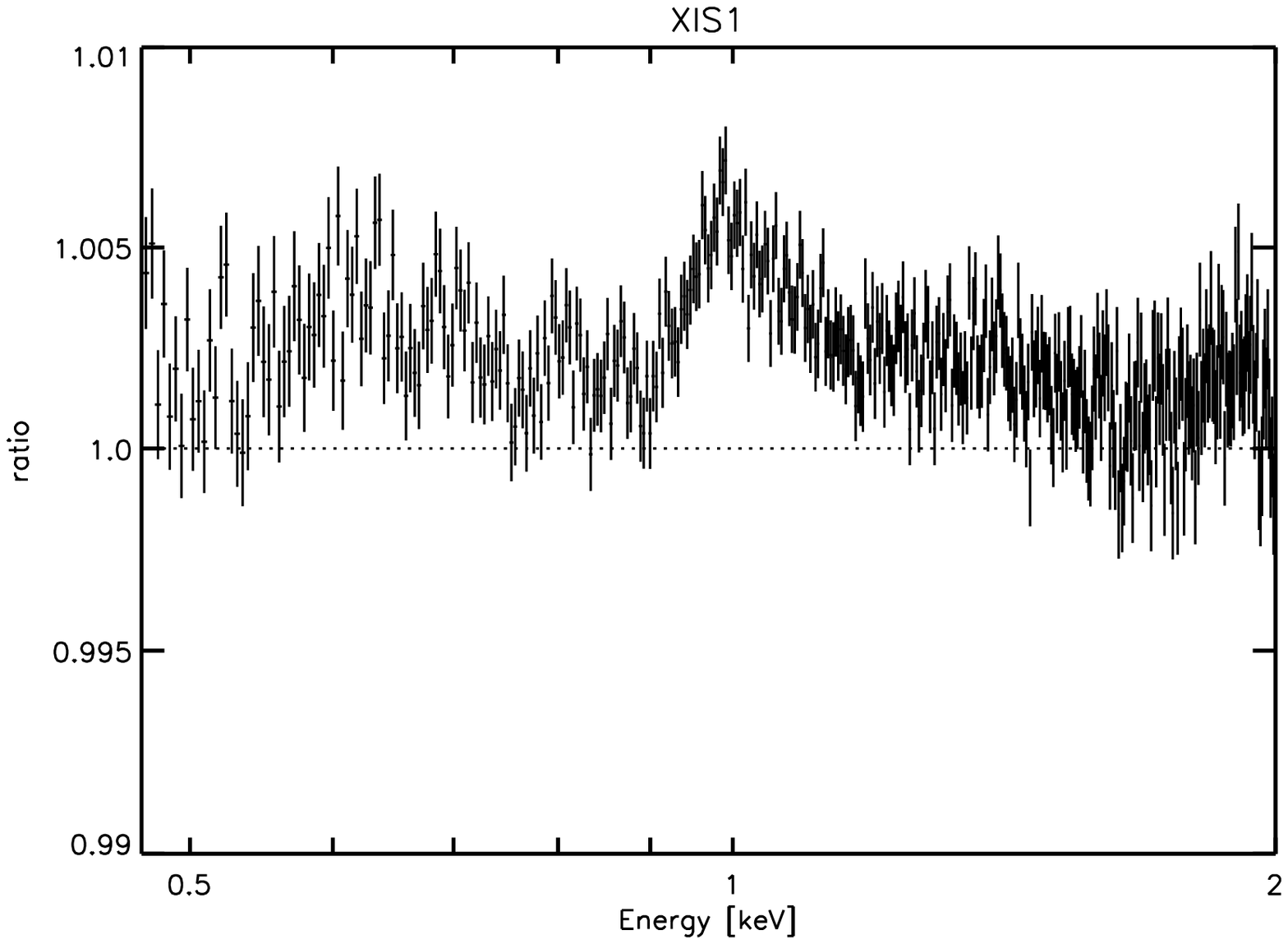}
\caption{The simulated A\,3112 XIS1 spectrum of the 3--6 arcmin region divided by the model prediction obtained from a fit
         to the 2.0--7.0 keV band data. Left panel: The simulated data contain the true emission from the 3--6 arcmin region and  
         the PSF-scattered flux from the central 0.5 arcmin region. Right panel: The simulated data contain the true emission from
         the 3--6 arcmin region and the lower temperature components projected in the line of sight. }
\label{fake_ratio.fig}
\end{figure*}

The simulations showed that 9\% of the flux originating from the cool central 0.5 arcmin region ends up in the 3--6 
arcmin annulus. Using the above surface brightness profile we found that the true cluster flux originating from the 3--6 
arcmin region is comparable to that originating from the central 0.5 arcmin region. Thus, $\sim$10\% of the observed flux
within the 3--6 arcmin annulus originates from the cool core due to PSF scatter.  

To examine the effect of the PSF-scattered component, we used XSPEC to simulate data of a two-temperature plasma where 
the temperatures are as explained above (5.5 keV and 3.5 keV), metal abundances of 0.3 and 1.0 Solar and relative 
model normalisations of 0.9 and 0.1. We then fitted the simulated spectrum with a single temperature MEKAL model with a 
varying low-energy cut-off, similarly as for the observational data (see Section \ref{cutoff}).
The resulting best-fit temperatures with any cut-off value are the same within 2 \%, i.e. the PSF scatter does not explain 
the cut-off effect. This is consistent with the XMM-Newton results, which are not contaminated by the PSF effect. 
For illustration, we extrapolated the E$>$2 keV band best-fit model towards lower energies 
(see Fig. \ref{fake_ratio.fig}). The extrapolated model is consistent with the data within $\sim$2\%  
except for the energies around 1 keV where the higher metal abundance from the core produces $\sim$4\% residuals, 
similarly as with the observational data (see Fig. \ref{extrap.fig}).

\subsection{Projection}
\label{proj_app}
One possible reason for the difference of the temperatures obtained in different bands is the projection of emission from 
the cooler gas in the line of sight due to the radial decline of the intracluster gas temperature. As shown in Bonamente et 
al. (2007), in the 1.0--2.5 arcmin region of A\,3112 the projection effect is negligible. Since the region in the current 
work is at larger radii, the fraction of projected cooler emission may be different. Using the method of the above work, 
we produced a composite spectrum from the different temperatures projected in the line of sight for the 3--6 arcmin 
regions. We distributed the emission measure using the double-beta model obtained in Section \ref{psf_app} and the 
temperature using an average Chandra profile (Vikhlinin et al., 2006) with $<T>$ = 5.0 keV). We then fitted the composite 
spectrum within the same bands as the observational data above, and found that the best-fit temperature remains constant. 
Thus the projected multi-temperature gas does not explain the difference in the measured temperatures. For illustration, 
we extrapolated the best-fit model from the 2.0-7.0 keV band towards lower energies. Comparison with the simulated data shows 
that  while there are some residuals in the soft band, their magnitude is only 0.5\% of the extrapolated model at maximum 
(see Fig. \ref{fake_ratio.fig}).

\section{Modeling the wide band emission}
Having verified above that the X-ray excess signal we detected at energies $E>0.4$ keV is not an artifact but rather a 
real celestial phenomenon, we attempted here to model the excess emission in the XIS1 unit either by an additional MEKAL 
model or by a power-law model. The data quality does not allow to constrain all the parameters simultaneously and thus we 
need to assume the values for some parameters, as follows.

In the non-thermal scenario, we fixed the photon number index $\alpha_{ph}$ to 1.5, 2.0 and 2.5, which is a reasonable range of values
assuming Inverse Compton scatter of the CMB photons off the relativistic cluster electrons accelerated by a merger 
shock (e.g. Sarazin, 1999). The best-fit models are acceptable ($\chi^2$: 255.8, 259.8, 260.2 with 311 degrees of freedom for $\alpha_{ph}$ 1.5, 2.0 and 2.5, 
respectively), 
and yield that the non-thermal component may constitute $\sim$15\% of the total 0.4--7 keV flux.

\begin{figure}
\resizebox{\hsize}{!}{\includegraphics[angle=-90]{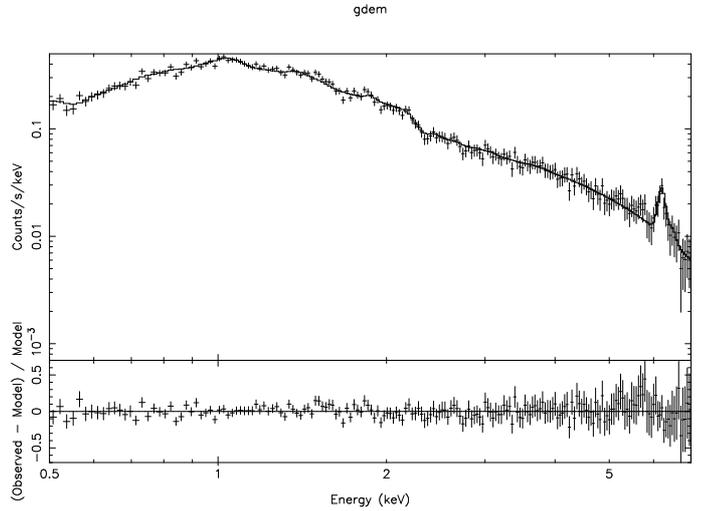}}
 \caption{The best-fit GDEM model for the XIS1 unit (upper panel) and fit residuals (lower panel).}
 \label{fig: gdem}
\end{figure}

The thermal model yields statistically equally good results as the non-thermal model.
The modeling is further complicated because the temperature and metal abundance of the additional component are strongly 
correlated.
Choosing a WHIM-like solution (e.g. Cen \& Ostriker, 1999) of abundance fixed to 0.1 and temperature to 0.1 keV yields 
large uncertainties in the derived emission measure, rendering it consistent with the WHIM simulations.
Assuming that the excess emission originates from material occupying the same volume as the hot gas, we forced the metal 
abundance of the additional component to be the same as that of the hot gas. 
The resulting large uncertainties in the temperature and emission measure allow 
clumpy lower temperature gas model of Cheng et al. (2005) as well as two phase 
gas occupying the same volume in pressure equilibrium.

\begin{figure}
\resizebox{\hsize}{!}{\includegraphics{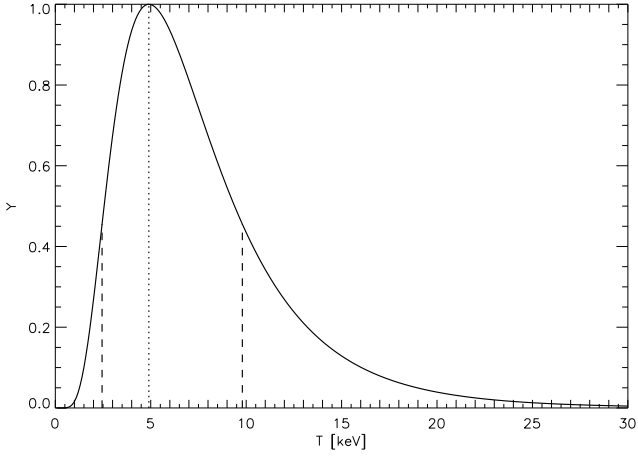}}
 \caption{The temperature--emission measure relation of the GDEM model. The dotted line shows the best-fit 
          temperature and the dashed lines show 0.5$\times$ and 2$\times$ the best-fit temperature. The emission measure is
          normalised to 1.0 at the best-fit temperature.}
 \label{fig: Y_gdem}
\end{figure}

\begin{table*}
 \caption{Single-temperature and multi-temperature XIS1 best-fit values.}
 \label{tab: dem}
 \begin{tabular}{l c c c c c c c}
 \hline\hline
  Model & $kT$\tablefootmark{a} & $kT_{\textrm{max}}$ & $\alpha$ & $\sigma_T$ & Abu\tablefootmark{b} & $\chi^2$ & d.o.f \\
        & [keV]  & [keV]               &          &       &         &          &       \\
 \hline
  CIE   & 4.7$\pm$0.1         &                         &                        &                        & 0.33$\pm$0.4 & 151.59 & 194 \\
  WDEM  & 4.9$_{-0.8}^{+1.1}$ & 6.2$\pm$0.7             & 0.36$\pm$0.18          &                        & 0.32$\pm$0.4 & 149.67 & 193 \\
  GDEM  & 4.9$\pm$0.2         &                         &                        & 0.24$_{-0.08}^{+0.06}$    & 0.32$\pm$0.4 & 148.15 & 193 \\
 \hline
 \end{tabular}
\begin{flushleft}
 \tablefoottext{a}{For the WDEM model $kT=kT_{\textrm{mean}}$ as shown in the Eq. \ref{eq: wdem T mean}.} 
 \tablefoottext{b}{Abundances are in solar units (Anders \& Grevesse 1989).}
\end{flushleft}
\end{table*}

We also attempted to model the excess emission in the XIS1 unit utilising two Differential Emission Measure (DEM) models, a 
Gaussian differential emission measure distribution (GDEM) (De Plaa et al., 2006) and WDEM (Kaastra et al., 2004) model.
For this multi-temperature modeling we used the SPEX package (Kaastra et al. 1996). 
Since the binning differs from that used in the XSPEC analysis we fitted the XIS1 data also using the SPEX's 
collisional ionisation equilibrium (CIE) model to make the DEM $\chi^2$ values and degrees of freedom (d.o.f) comparable to the 
single-temperature values (see Table \ref{tab: dem} and Fig. \ref{fig: gdem}). 
The hypothesis that the GDEM and WDEM model does not improve over the CIE model is rejected with a significance level of 3.6\%
and 11.7\%, respectively.

In the WDEM model the emission measure, $Y = \int n_{\textrm{e}}n_{\textrm{H}}$dV, of a number of thermal components is 
distributed as a truncated power law: 

\begin{equation}
 \label{eq: wdem}
 \frac{dY}{dT} = \left\{ \begin{array}{ll}
                 cT^{1/\alpha} & \verb|    | \beta T_{\textrm{max}} \leq T < T_{\textrm{max}} \\
                 0 & \verb|    | T > T_{\textrm{max}} \verb| | \textrm{or} \verb| | T < \beta T _{\textrm{max}}. \\
                 \end{array} \right.
\end{equation}
We adopted a value of 0.1 for $\beta$ from De Plaa et al. (2006). The best-fit $T_{\textrm{max}}$ and $\alpha$ parameters are 
6.2$\pm$0.7 keV and 0.36$\pm$0.18, respectively. 
In order to compare the temperature of the WDEM model with single-temperature models, we have calculated the 
mean temperature for the WDEM model as reported in De Plaa et al. (2006):

\begin{equation}
 \label{eq: wdem T mean}
 T_{\textrm{mean}} = \frac{(1+1/\alpha)}{(2+1/\alpha)}\frac{(1-\beta^{1/\alpha+2})}{(1-\beta^{1/\alpha+1})}T_{\textrm{max}}.
\end{equation}
The best-fit $T_{\textrm{mean}}$ is 4.9$_{-0.8}^{+1.1}$ keV.

In the GDEM model the emission measure distribution is Gaussian:

\begin{equation}
 \label{eq: gdem}
 Y(x) = \frac{Y_0}{\sigma _T \sqrt{2\pi}}e^{-(x-x_0)^2/2\sigma _T^2},
\end{equation}  
where $x$~=~log~$T$ and $x_0$~=~log~$T_0$. $T_0$ is the average temperature of the distribution and $\sigma_T$ is 
the width of the Gaussian. The best-fit temperature and $\sigma_T$ parameter values are 4.9$\pm$0.2 keV and 0.24$_{-0.08}^{+0.06}$, respectively.
At the temperature of 0.5$\times$ and 2$\times$ the best fit temperature the emission measure is dropped to 46\% of the maximum (Fig. \ref{fig: Y_gdem}).

The temperatures of the DEM modeling are consistent with the single-temperature model value (see Table \ref{tab: dem}). The $\alpha$ and $\sigma_T$ 
values deviate from the zero by a 2 $\sigma$ and 3 $\sigma$ confidence levels, respectively. This indicates that the intra-cluster
plasma consists of multi-temperature plasma. The temperature variations may be on different spatial scales within the extraction region of the spectrum. 
Such differences may easily arise due to the history of the cluster, which every now and then accretes other clusters or groups. The gas has therefore different origins and
therefore different temperatures and densities, and although pressure equilibrium will be established rather soon, to wash out the temperature
fluctuations takes much longer especially given the long cooling time of the hot gas.
The parameter values of the DEM modeling are similar to the values found in the studies of  
Kaastra et al. (2004) for several clusters and De Plaa et al. (2006) for S\'ersic 159-03.

\section{Conclusions}
We have analysed Suzaku XIS1 data of A\,3112 cluster of galaxies. We found that in the 3--6 arcmin annulus around the cluster
center, the best-fit temperature depends strongly on the choice of the energy band used for the spectral analysis:
the temperature changes from 5.7 keV to 4.7 keV when varying the low energy cut-off from 2.5 keV to 0.4 keV.
We showed that this effect is not due to uncertainties in the modeling of the background or the Galactic absorption, 
PSF scatter from the bright cool core, or line-of-sight projection of emission from cooler gas at large radii. Also, a 
calibration effect is not likely, since a qualitatively similar effect is present in the XMM-Newton data of A\,3112. Thus, 
the strong dependence of the temperature on the low energy cut-off yields evidence for excess X-ray emission in A\,3112.  

Assuming that the emission above photon energies of 2 keV originates purely from the hot intracluster gas, 
the excess emission is $\sim$10\% of the cluster hot gas emission at 0.4--1.7 keV. 
Using the old Galactic NH (Dickey \& Lockman, 1990) the excess emission level is twice as high as in the Kalberla et al. (2005)
at 0.6 keV and almost thrice as high at 0.4 keV,
while at higher energies the effect is not significant compared to the statistical uncertainties.

Due to the small differencies between thermal and non-thermal model we can not rule out the non-thermal origin of the 
excess emission based on the goodness of the fit. Assuming that it has a thermal origin, we
further examined the DEM models. Using the GDEM or WDEM model improves the fit compared to the single temperature model. 
The deviations from the single-temperature phase, as measured in terms of WDEM and GDEM parameters, are significant at 2 $\sigma$ and 
3 $\sigma$ level, respectively.

\begin{acknowledgements}
 The work is based on observations obtained with {\it Suzaku} and {\it XMM-Newton}
 satellites. TL and JN are supported by the Academy of Finland.
\end{acknowledgements}

\begin{appendix}

\section{XMM-Newton data reduction}
\label{xmm_app}
We analysed the observation of A\,3112 with observation number 0105660101, performed on Dec 24, 2000 (the same set as in 
Nevalainen et al., 2004). We used SAS version 9.0.0. for the reduction and response generation, using the latest 
calibration information available on Nov 2009.
The raw data were processed with epchain version 8.63 to produce the event files. Additionally, for PN we produced the
out-of-time event file which we used to correct the PN spectra. 
We filtered the data using the SAS expression ``flag==0'', which rejects the data from bad pixels and CCD gap regions.
We further filtered the events by allowing patterns 0-4 for PN and 0-12 for MOS.

The E $>$ 10 keV light curve was very flat, indicating that the data were not significantly contaminated by the 
particle flares. The useful exposure time was 16 ks (PN) and 45 ks (combined MOS1~+~MOS2).

For the sky background component, we used the same method as for the Suzaku data (see Section \ref{sky}) i.e.
the RASS spectrum around the cluster.
We used the closed filter data sets from Nevalainen et al. (2005) to extract the particle-induced background spectra
in the same detector regions as used for the cluster data. After scaling the closed filter spectra by the E $>$ 10 keV 
count rates, i.e by a factor of 1.11 and 1.01 for PN and MOS, we subtracted these from the cluster data.

\end{appendix}

\end{document}